\documentclass[aps,prb,twocolumn,showpacs,a4paper,floatfix]{revtex4-1}
\usepackage{times}
\usepackage[latin1]{inputenc}
\usepackage[english]{babel}
\usepackage[T1]{fontenc}
\usepackage{graphicx}
\usepackage{subfig}
\usepackage{epsfig}
\usepackage{varioref}%
\input epsf
\usepackage{rotating}
\usepackage{mathptmx}
\usepackage{fancyhdr}
\usepackage{amsmath}
\usepackage{amsfonts}
\usepackage{amssymb}
\usepackage{amsbsy}
\usepackage{textcomp}
\usepackage{array}
\usepackage{fixmath}
\usepackage{bm}
\DeclareGraphicsExtensions{.pdf,.png,.jpg,.mps,.eps}
\usepackage[colorlinks]{hyperref}
\usepackage{eso-pic}

\begin{document}
\title{Electronic structure of fluorides: general trends for ground and excited state properties}
\author{Emiliano Cadelano}
\email[E-mail me at:]{emiliano.cadelano@dsf.unica.it}
\author{Giancarlo Cappellini}
\affiliation{Department of Physics, University of Cagliari and\\
Istituto Officina dei Materiali (IOM) del Consiglio Nazionale delle
Ricerche (CNR), Unita' Operativa SLACS
\\ S.P. Monserrato-Sestu Km 0.700, I-09042 Monserrato (Cagliari), Italy}
\date{\today}

\begin{abstract}
The electronic structure of fluorite crystals are studied by means of density functional theory within the
local density approximation for the exchange correlation energy.
The ground-state electronic properties, which have been calculated for the cubic structures $CaF_{2}$,$SrF_{2}$, $BaF_{2}$, $CdF_{2}$, $HgF_{2}$, $\beta $-$PbF_{2}$, using a plane waves  expansion of the wave functions, show good comparison with existing experimental data and previous theoretical results.
The electronic density of states at the gap region for all the compounds and their energy-band structure have been calculated and compared with the existing data in the literature.
General trends for the ground-state parameters, the electronic energy-bands  and transition energies for all the fluorides considered are given and discussed in details. Moreover, for the first time  results for $HgF_{2}$ have been presented.
\end{abstract}
\maketitle
\section{Introduction}
\label{sec:introdution}
Fluorides and fluorite-type crystals have attracted much interest for their intrinsic optical properties and their potential applications in optoelectronic devices. The technological importance of $CaF_{2}$ is due mainly to its optical properties; indeed $CaF_{2}$ has a direct band gap  at $\Gamma$ of $12.1$ eV ~and an indirect band gap estimated around $11.8$ eV.\cite{rubloff} ~Calcium fluorite $CaF_{2}$ as well as all the other fluorides, is a highly ionic insulator with a large band gap, and its lattice structure is a cubic \textit{ $Fm\overline{3}m$ } within three ions per unit cell, i.e. one cation placed in the origin and two anions $F$ are situated at $\pm(\frac{1}{4}a, \frac{1}{4}a,\frac{1}{4}a)$.\cite{wyckoff}
Here we shall consider $CaF_{2}$, $SrF_{2}$, $BaF_{2}$ (with cations belonging to the II group) and $CdF_{2}$, $HgF_{2}$ (with cation belonging to group IIB) and, finally, $\beta $-$PbF_{2}$ (with cation belonging IV group). We refer to them respectivelly in the present text as II-compounds, IIB-compounds and $\beta $-$PbF_{2}$.
\begin{figure}[b]
\centering%
\subfloat[conventional cell\label{capt:structure}]%
{\label{fig:structure}\includegraphics[width= 0.22\textwidth]{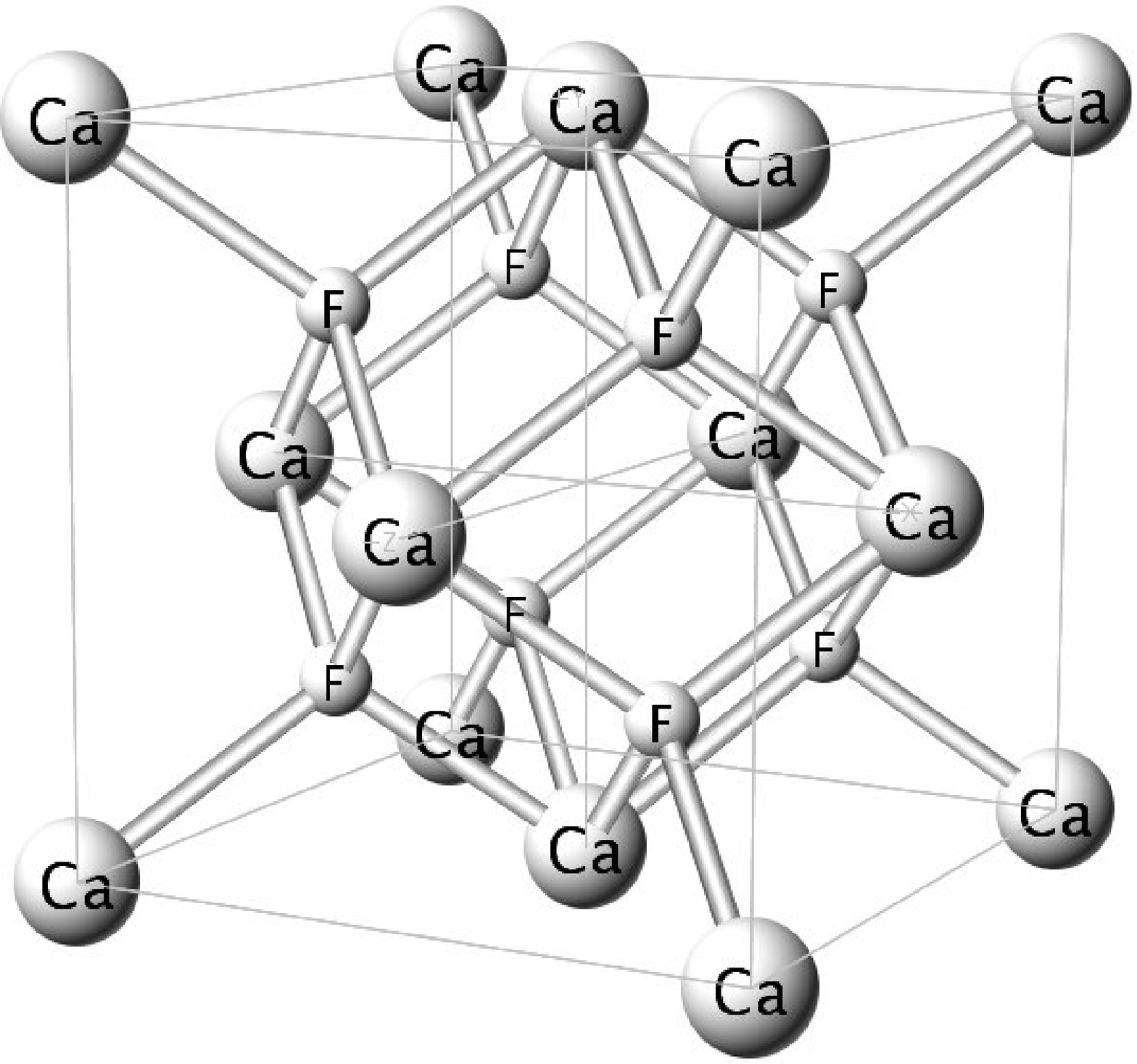}}
\subfloat[Brillouin zone for $Fm\overline{3}m$\label{capt:bz}]%
{\label{fig:bz}\includegraphics[width= 0.22\textwidth]{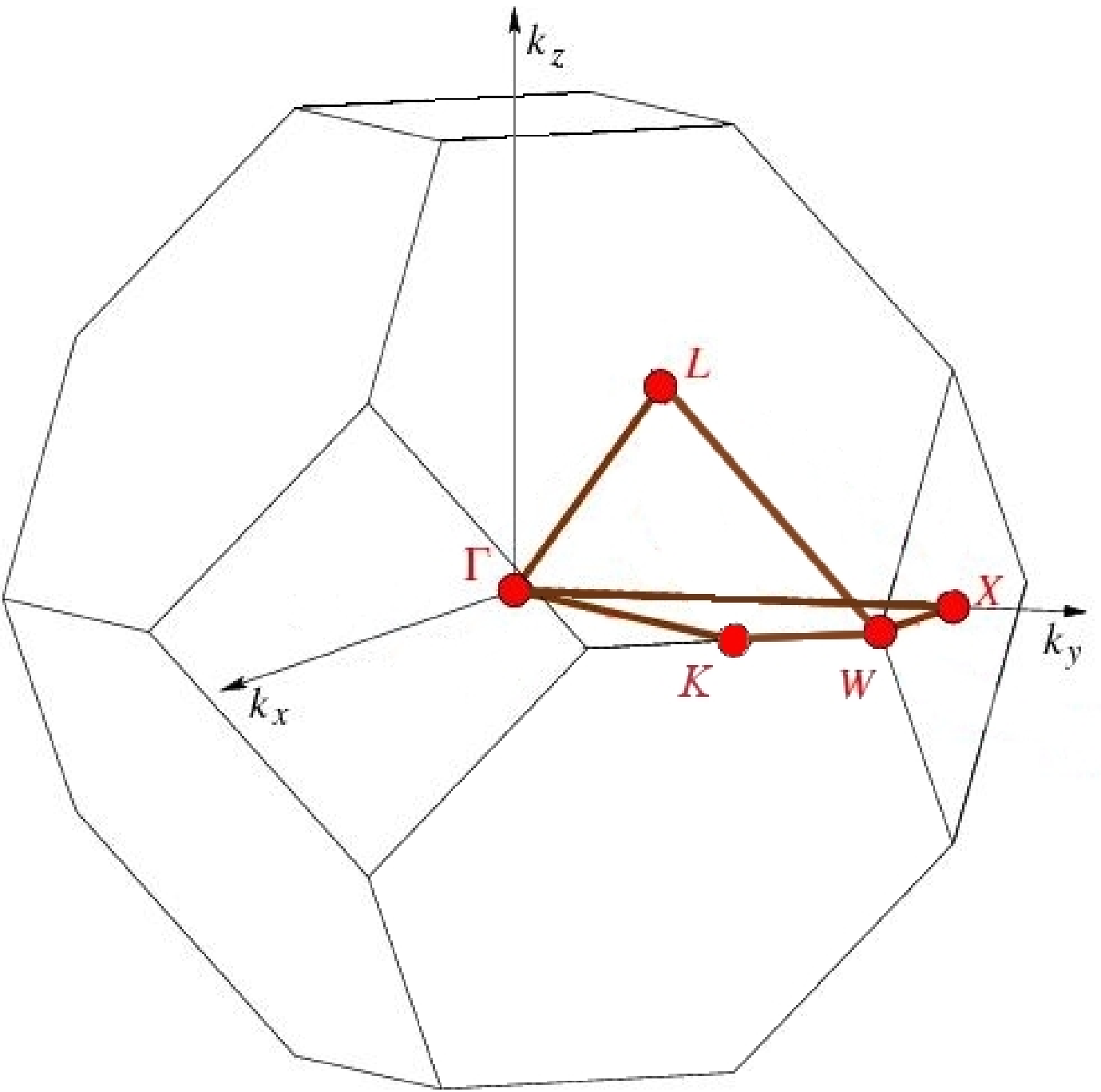}}%
\caption{\label{fig:structure+bz}(a) Fluorite structure. Small spheres represent $F^{-}$ ions and the larger ones the cations$^{++}$. 
(b) The Brillouin zone of space group \textit{ $Fm\overline{3}m$}.}
\end{figure}
We therefore propose the first systematic study of the electronic properties of some fluorides compounds within the same computational approach. Until now the theoretical studies of these compounds have been tackled in separate forms and with  different techniques. Therefore no general and  systematic trend for the family of these compounds could be obtained. The DFT-LDA studies are of particular importance and are benchmarks for subsequent researches to perform excited states and optical properties calculations.  The latter rely on more sophisticated techniques which must start from well converged DFT-LDA calculations (e.g. perturbative $G_{o}W_{o}$, self-consistent GW, BSE etc.).\cite{cappellini1, cappellini2} 
In the following lines we proceed to consider the experimental data and the theoretical results for fluorides present in the literature.
\paragraph{Experimental Scenario}
Due to its importance in application and basic research, experiments on $CaF_{2}$ and fluorides compounds have been carried out for a long time.
The optical constant of $CaF_{2}$ in the extreme ultraviolet has been studied by discharge-tube technique.\cite{tousey:1936}
Reflectance spectra, transition energies for  $CaF_{2}$, $BaF_{2}$, $SrF_{2}$ and other ionic compounds  have been measured by synchrotron radiation facility later on.\cite{rubloff} Studies on 
$\beta $-$PbF_{2}$, $CaF_{2}$, $SrF_{2}$ e $BaF_{2}$ dielectric properties have been published  as a function of pressure and temperature by capacitance and dielectric loss measurements.\cite{Samara:1976}
Satellites in the $X$-ray spectra for $CaF_{2}$, $SrF_{2}$, $BaF_{2}$ , density of states for intraband transitions, 
have been studied by photoelectron spectrometry.\cite{Scrocco}
The effects of Eu defects in $\beta $-$PbF_{2}$ ($PbF_{2}:Eu^{3+}$) relative to fluorescence/electronic excitation spectra and dielectric relaxation have been analyzed  by laser absorption spectra techniques.\cite{weesner:1986}
$\beta $-$PbF_{2}$ and $CdF_{2}$ mixed crystals  absorption coefficients have been reported
by spectrophotometry measurements.\cite{kosacki:1986}
Neutron diffraction techniques have been employed to determine the $\beta $-$PbF_{2}$ pressure and temperature  phase diagram.\cite{hull}
$\alpha - PbF_{2}$, $\beta $-$PbF_{2}$ and others compounds absorption spectra and electronic transitions 
have been studied by polarized reflectivity.\cite{Fujita}
Polarized deep and vacuum ultraviolet light measurements permitted the study of birefringence of $CaF_{2}$ and $BaF_{2}$.\cite{burnett} 
The study by different techniques of the phase diagram of $HgF_{2}$ and other Hg compounds  should also be mentioned here \cite{hostettler:2005} (even if for $HgF_{2}$ no results appear in this work due to hydration of the sample). 
Recently schematic experimental phase diagrams for $HgF_{2}$ and $HgF$ have been also reported.\cite{okamoto:2008}
\paragraph{Theoretical Scenario} 
Various  theoretical methods have been applied to study either the ground state or the excited states of the fluorides compounds.
One of the first works of relevance is the one in which elastic constants, pressure derivatives of $2^{nd}$ order elastic constants, static dielectric constant and its strain dependence  for  $CaF_{2}$, $SrF_{2}$, $BaF_{2}$, have been calculated within a shell model.\cite{Srinivasan}
The energy bands and reflectance spectra of $CaF_{2}$ and $CdF_{2}$  have been determined afterwards
within  a combined tight-binding and pseudopotential method.\cite{albert:1977}
Mixed crystals of $CaF_{2}$, $SrF_{2}$, $CdF_{2}$, $\beta $-$PbF_{2}$ have been studied with respect to their 
energy bands and DOS  within  LMTO method.\cite{kudrnovsky}
A phenomenological method has been then applied to calculate 
specific heat of $\beta $-$PbF_{2}$, $CaF_{2}$ and other compounds.\cite{bouteiller:1992}
Linear and non-linear optical properties of the cubic insulators $CaF_{2}$, $SrF_{2}$, $CdF_{2}$, $BaF_{2}$ and other compounds  have been determined by first principles OLCAO.\cite{ching}
Points defects study in $CdF_{2}$ have been performed within the plane wave pseudopotential method.\cite{Mattila} With respect to electronic excitation properties and energy band-gaps, electronic band structure of $CaF_{2}$ and other compounds have been determined by DFT-GW, using plane wave basis set and ionic pseudopotentials (PW-PP) scheme.\cite{shirley}  
Using the hybrid B3PW functional, the electronic structures of defected fluorides, namely  $CaF_{2}$ and $BaF_{2}$, have been calculated. \cite{ranjia,ShiEglitis,ShiEglitis2} 
The $\varepsilon_{2}(\omega)$ function after an iterative procedure using an effective Hamiltonian has been calculated for $CaF_{2}$ and GaN,\cite{benedict:1999} within PW-PP considering a  screened interaction for \textit{e-h} coupling. Native and rare-earth doped defects complexes in $\beta $-$PbF_{2}$ have been studied by atomistic simulation.\cite{jang:2000}

In this paper, we are interested in either the structural and the electronic properties of each fluorite, and in comparison/trend studies for the whole crystallographic family.
We have computed therefore the electronic and structural properties of different fluorides, $CaF_{2}$, $SrF_{2}$, $BaF_{2}$ (with cation belonging to group II), $CdF_{2}$, $HgF_{2}$ (with cation belonging to group IIB), $\beta $-$PbF_{2}$, within the same first-principles pseudopotential plane-wave method.

\begin{table}[tp]
\caption{Optimized lattice constants of fluorides.  In columns ''LDA'' and ''Theory`` we show previuos theoretical results (DFT-LDA and others respectively), while in column  ''Exp.'' data after different experimental references are reported.}
\label{lattice constant}
\centering
 \begin{tabular}{c|c|ccc}
\hline
\hline
$a_{o}$[\AA] &  { present }     & LDA  & Theory &  Exp.\\
\hline
$CaF_{2}$   		&   {5.30}  &  5.34\cite{Kalugin}& 5.46\cite{kudrnovsky} 	& 5.46\cite{Weast} \\
$SrF_{2}$   		&   {5.68}  &   -              & 5.79\cite{kudrnovsky}		& 5.78\cite{Srinivasan}	\\
$BaF_{2}$   		&   {6.09}  &   -	       & 6.26\cite{ranjia}		& 6.17\cite{Srinivasan}	\\
$CdF_{2}$   		&   {5.31}  &  5.36\cite{Deligoz} & 5.39\cite{kudrnovsky}	& 5.46\cite{Kalugin}\\
$HgF_{2}$   		&   {5.47}  &   -	       & 5.55\cite{kaupp} 		& 5.54\cite{Ebert}	\\
$\beta $-$ PbF_{2}$     &    5.77   &   - 	       & 5.94\cite{kudrnovsky} 		& 5.94\cite{jang:2000}	\\
\hline
\hline
\end{tabular}
\end{table}

\section{Computational details}
\label{sec:method}
All the calculations for the fluorides under study have been performed using density funcional theory (DFT)\cite{kohn} method implemented in the plane-wave basis code VASP.\cite{Kresse,Kresse2} Projector augmented wave pseudopotentials (PAW)\cite{paw,paw2} have been used in the localized density approximation (LDA) for the exchange correlation energy treated within the scheme of Ceperley and Alder parametrized by Perdew and Zunger.\cite{Cerperley,pervew}

Relativistic effects have been included in the calculations\cite{Kresse,Kresse2, bachelet, kaupp} and spin-orbit coupling has been considered for the valence electrons.\cite{koelling}

The conventional cell of the crystals is shown in Fig.~\ref{capt:structure}, in which the ions $F^{(-)}$ (drawn as little spheres) form a cubic sublattice surrounded by a faced cubic center lattice of cations$^{(++)}$ (in the picture shown as large spheres labeled e.g. as Calcium).
Fluorides with cations belonging to the II and IIB groups show a stable phase in this crystallographic structure.
On the other hand, at low pressure the $PbF_{2}$ show two structural phases, namely orthorhombic ($\alpha$) and cubic ($\beta$). Although the cubic phase  $\beta $-$PbF_{2}$ is the most stable in ambient condictions, while the orthorhombic $\alpha $-$PbF_{2}$ becomes stable at high pressure.\cite{jiang2,nizam}

In Fig.\ref{capt:bz}, the Brillouin zone has been drawn for the \textit{ $Fm\overline{3}m$ } space group; the paths in the $k$-space chosen in our calculations for the electronic band structures have also been shown.
All calculations are performed with the total energy convergence within $1.5\cdot 10^{-5}{}$eV  with kinetic energy cut-off depending on the cation of the compound under study (at least of $550 {eV}$).
A Monkhort-Park\cite{monkhort} $k$-point mesh of $4\times4\times4$ has been chosen to relax the cell parameters, till the largest value for the interatomic forces result smaller than  $1.5\cdot 10^{-5}{}$ eV/\AA.
Lattice constants for each fluorite ($a_{\circ}$ in Tab.~\ref{lattice constant}), as well as the bulk moduli  ($B_{\circ}$ in Tab.~\ref{bulk moduli}) have been  calculated with the Murnaghan equation of state.\cite{murnaghan}
\begin{table}[tp]
\caption{Bulk moduli of fluorides. Data in column  ''Theory'' are previous theoretical results. Data in column ''Exp.'' are after different experimental references.}
\label{bulk moduli}
\centering
 \begin{tabular}{c|c|cc}
\hline
\hline
$B_{\circ}$[GPa] &  present  & Theory   &  Exp.\\
\hline
$CaF_{2}$  		 &        103.01      &  91\cite{sun}     \ &   90-82\cite{sun}     \\
$SrF_{2}$  		 &         83.75      &  -  		  \ &      -			\\
$BaF_{2}$  		 &         69.39      &  50\cite{Ayala}   \ &  59\cite{wong}    		\\
$CdF_{2}$   		 &        123.96      &  123\cite{Deligoz}\ &      -			\\
$HgF_{2}$  		 &        117.03      &  -  		  \ &     -			\\
$\beta $-$PbF_{2}$  	&  	   93.22      &  60\cite{jiang2}  \ &  64\cite{hull}   		\\
\hline
\hline
\end{tabular}
\end{table}
\section{Results and discussion}
\label{sec: result}
\begin{figure}[bp]
\centering%
\includegraphics[width= 0.48\textwidth]{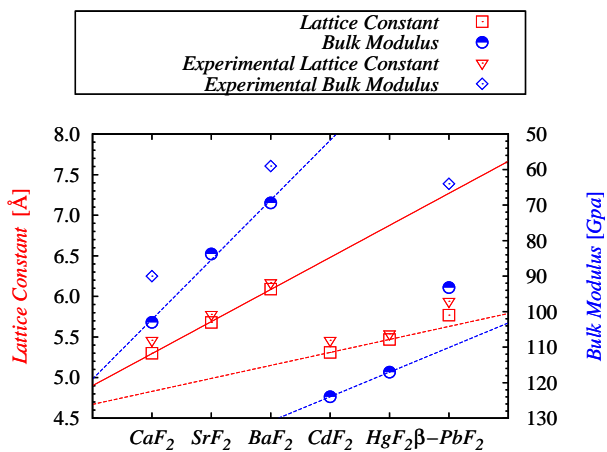}
\caption{Lattice constant and bulk modulus for each fluorite under study.  Results after present study Vs. available experimental data.}
\label{trendbulk}
\end{figure}

\begin{table*}[tbp]
\caption{For the fluorides considered here, bandwidths and energy gaps as found in our DFT-LDA study. Direct and indirect DFT-LDA band gap energies. In the first column, current results are shown. Data are in eV.}
\label{tab:gap bandwidht}
\centering
 \begin{tabular}{l|ccccc|ccccc|cccc}
\hline
\hline
&\multicolumn{5}{c}{Minimum Direct Band gap} &\multicolumn{5}{c}{Minimum Indirect Band gap} &\multicolumn{4}{c}{Valence Bandwidth} \\
Solid &   &  { present }  & LDA & GW &  Exp. & & { present }  & LDA & GW &  Exp.  & { present }  & LDA & GW &  Exp.  \\
\hline
$CaF_{2}$ & $ \Gamma \rightarrow \Gamma $  &  {7.71}   &  7.11\cite{rohlfing}  & 11.80\cite{rohlfing} & 12.10\cite{rubloff} &  $ X\rightarrow \Gamma $  &  {7.43}   &  6.85\cite{rohlfing} & 11.50\cite{rohlfing} & 11.80\cite{rubloff} & {3.17}	&	2.82\cite{rohlfing}&	3.26\cite{rohlfing} &	3.20\cite{Scrocco}\\
$SrF_{2}$& $ \Gamma\rightarrow \Gamma $  &  {6.99}      &	&	& 11.25\cite{rubloff} &
 $ X\rightarrow \Gamma $ &  {6.89}	& 6.77\cite{ching}& & 10.60\cite{rubloff}	&{2.33}	&	&	  & 2.80\cite{Scrocco} \\
$BaF_{2}$&  $ \Gamma\rightarrow \Gamma $ &  {6.67}  	 &	&	& 11.00\cite{rubloff}	& 
$(\tfrac{1}{4},~\tfrac{1}{4},~0) \rightarrow\Gamma$	& {6.58} & 7.19\cite{ching}	&	&10.00\cite{rubloff}	&
{1.78}	& 	&	& 2.50\cite{Scrocco}\\
$CdF_{2}$&  $ \Gamma\rightarrow \Gamma $ &   {3.37} 	 & 3.30\cite{Mattila}	&	& 	&$W\rightarrow\Gamma$&
{2.94} &	 2.80\cite{Mattila}  & 	& 7.80\cite{Orlowski}	&  5.79	&	&	&\\
$HgF_{2}$&  $ \Gamma\rightarrow \Gamma $ &   0.35	 &	&	& 	& 
$ \Gamma\rightarrow L $	& 4.16 & 	&	&	&
6.38	& 	&	& \\
$\beta $-$PbF_{2}$& $ X\rightarrow X $  &   {4.09}	 & 	&	&	
& $ W\rightarrow X $  & {3.41}	&            &        &         &{7.16}	&	&  & 6.33\cite{Fujita}	\\
\hline
\hline
\end{tabular}
\end{table*}
Trends for lattice constants and bulk moduli are showed in Fig.~\ref{trendbulk} with the corresponding experimental data from literature, if available.
From Table~\ref{lattice constant} and Fig.~\ref{trendbulk} an overall good comparison appears between experimental data and present results relative to lattice constants with maximum deviation of $3\%$. Compounds with cations belonging to the II and IIB groups show a linear behavior with respect to Mendeleev table period of corresponding cation.
Moreover, as shown in Table \ref{bulk moduli} and Fig.~\ref{trendbulk}, bulk moduli show a less satisfactory agreement  with experimental results ( maximum deviation within $20\%$ in the case of II group, within $45\%$ for the $\beta $-$PbF_{2}$). Also for bulk moduli of compounds with cations belonging to the II and IIB groups, an almost linear behavior results with respect to the Mendeleev's period.
For IIB group no available experimental data supports the theoretical trend.
\begin{figure}[bp]
\centering
\includegraphics[width= 0.48\textwidth]{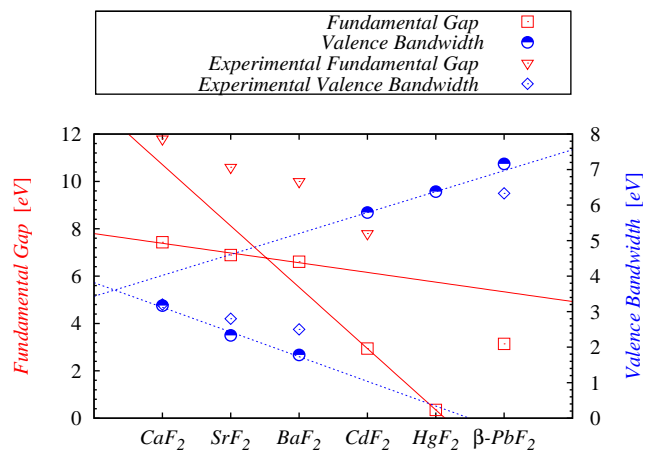}
 \caption{Minimum energy gaps and valence bandwidths trends for the compounds under study. Also reported available experimental data.}
\label{trendenergy}
\end{figure}

\begin{figure}[tb]
\centering%
\subfloat[$CaF_{2}$\label{fig:band ca}]%
{\label{figs:band ca}\includegraphics[width= 0.4\textwidth]{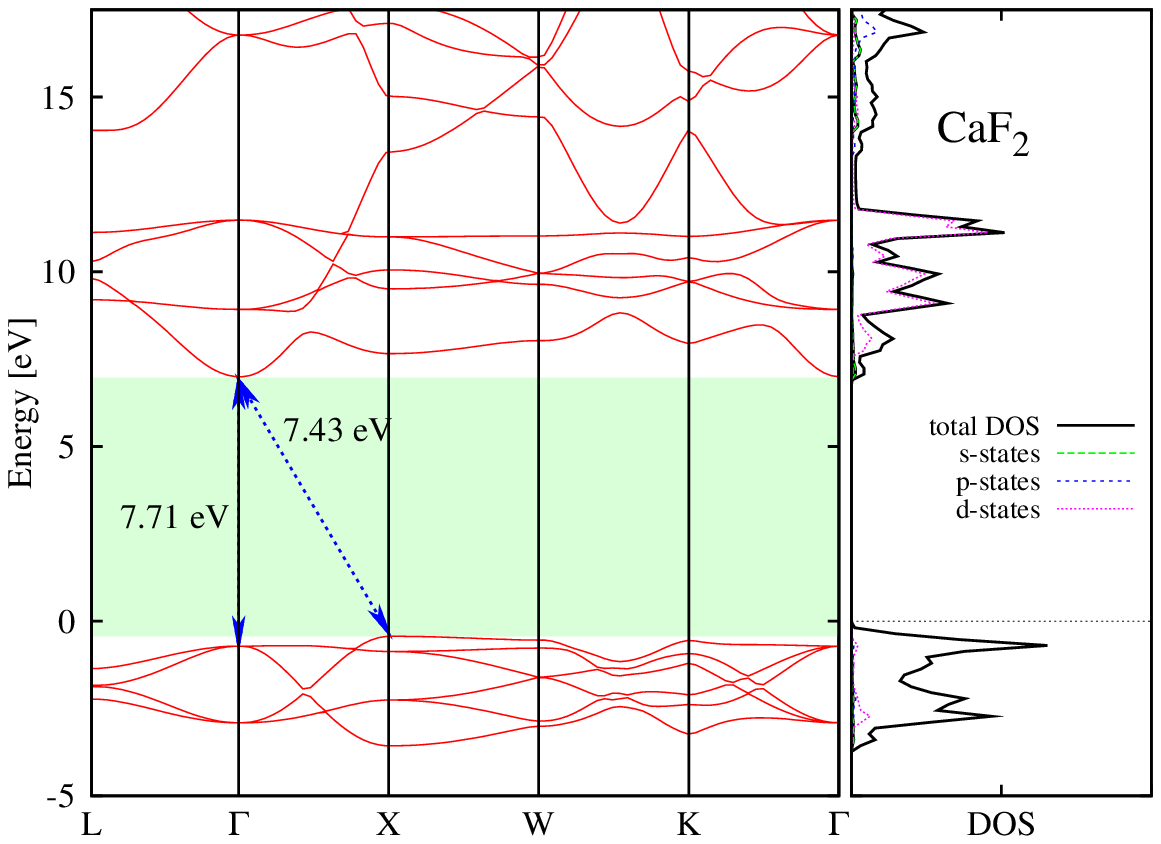}}\qquad
\subfloat[$SrF_{2}$\label{fig:band sr}]%
{\label{figs:band sr}\includegraphics[width= 0.4\textwidth]{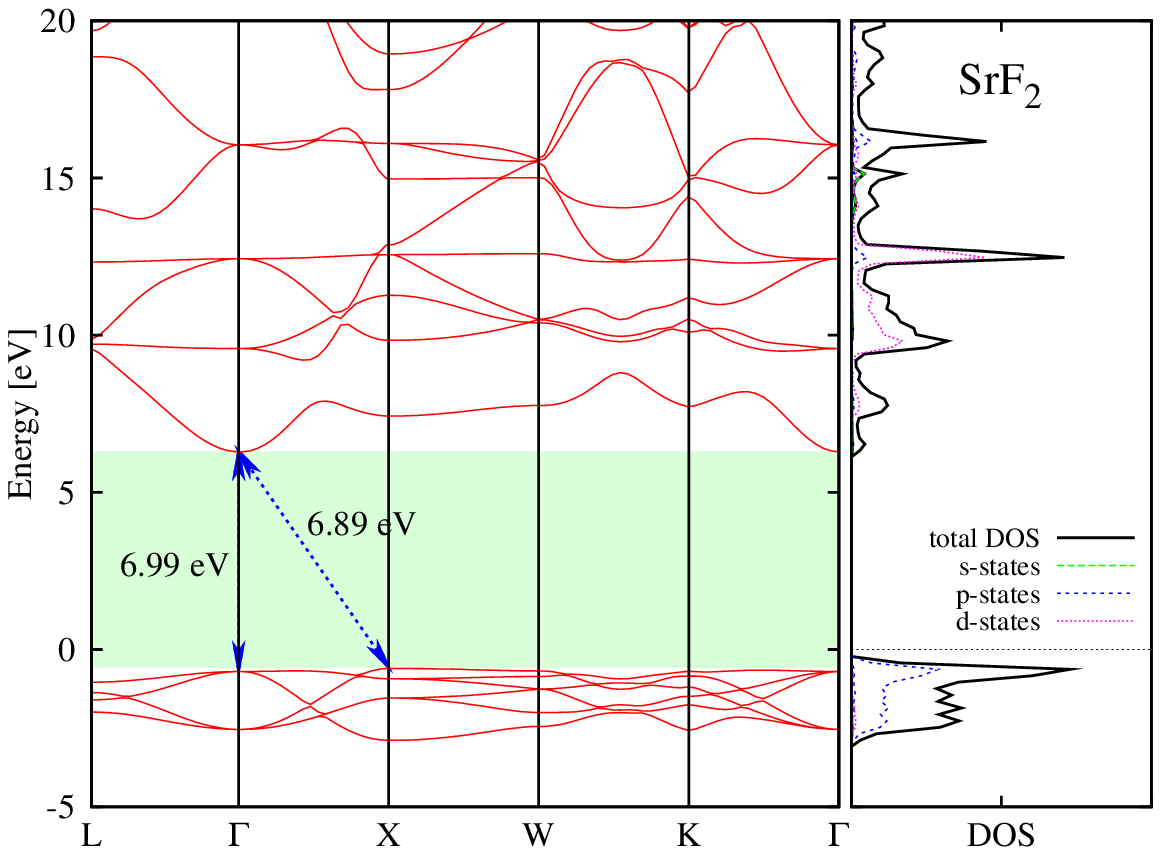}}\qquad
\subfloat[$BaF_{2}$\label{fig:band ba}]%
{\label{figs:band ba}\includegraphics[width= 0.4\textwidth]{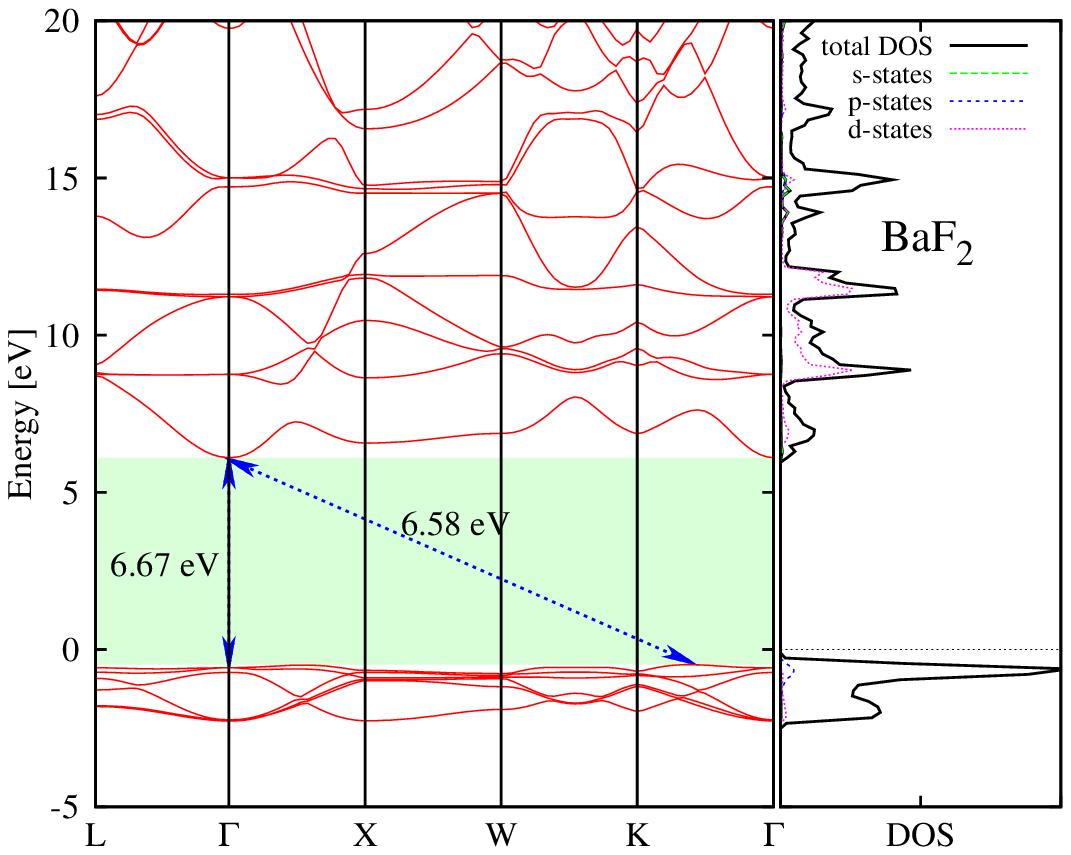}}
\caption{\label{cpt:band graph}Energy band structures, total and projected density of states (DOS) are shown in the region of the band gap for the II-compounds.}
\end{figure}

\begin{figure}[tb]
\centering%
\subfloat[$CdF_{2}$\label{fig:band cd}]%
{\label{figs:band cd}\includegraphics[width= 0.4\textwidth]{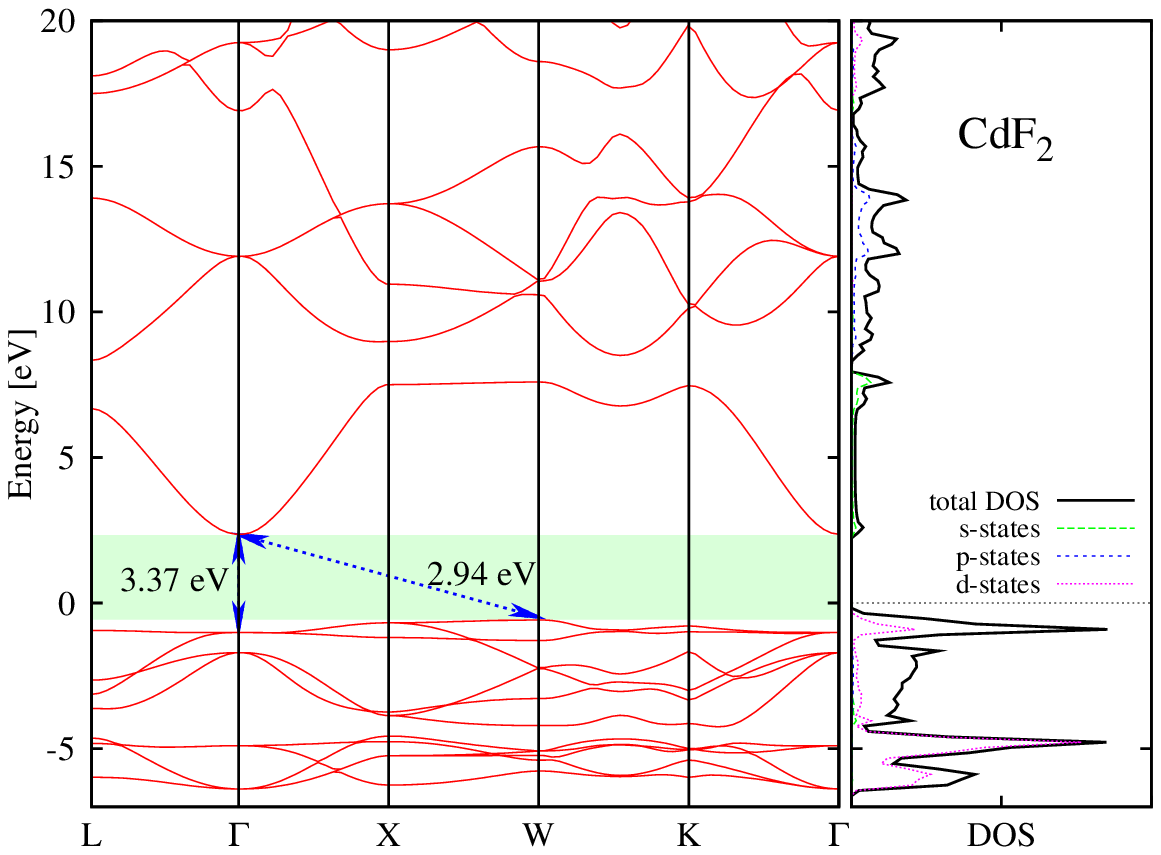}}\qquad
\subfloat[$HgF_{2}$\label{fig:band hg}]%
{\label{figs:band hg}\includegraphics[width= 0.4\textwidth]{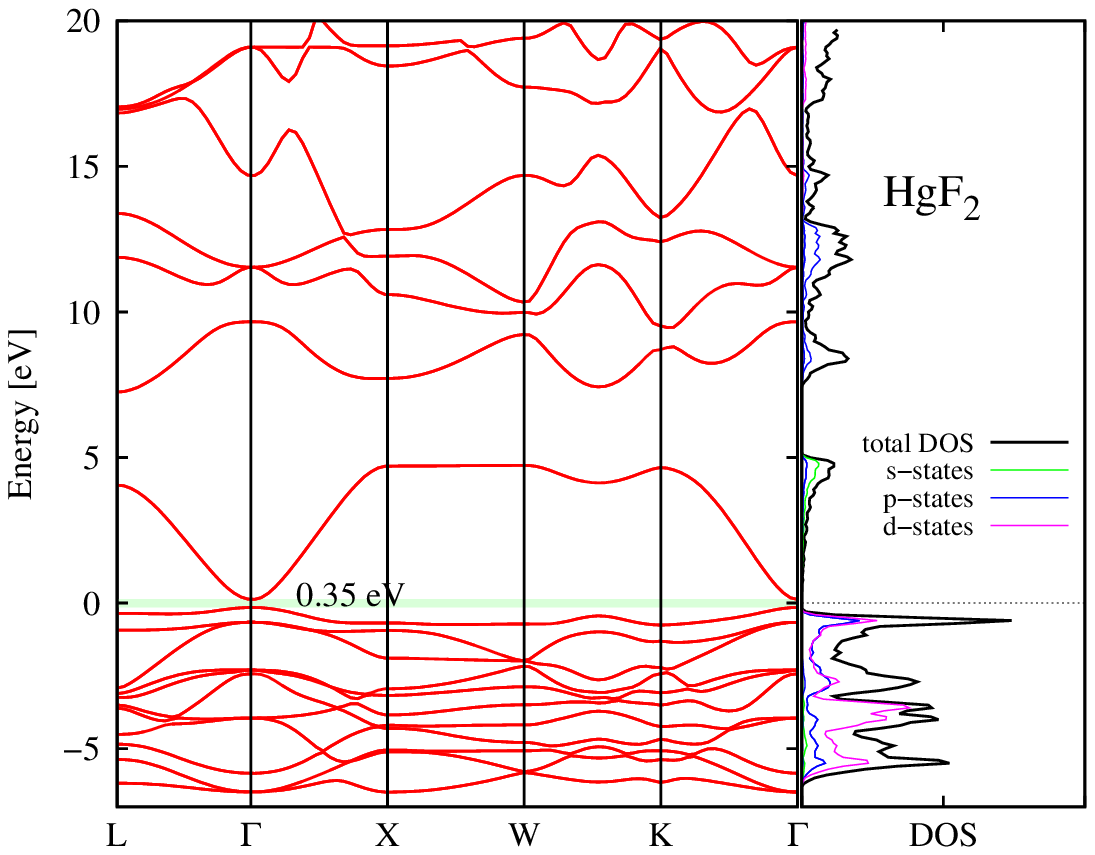}}\qquad
\subfloat[$\beta $-$PbF_{2}$\label{fig:band pb}]%
{\label{figs:band pb}\includegraphics[width= 0.4\textwidth]{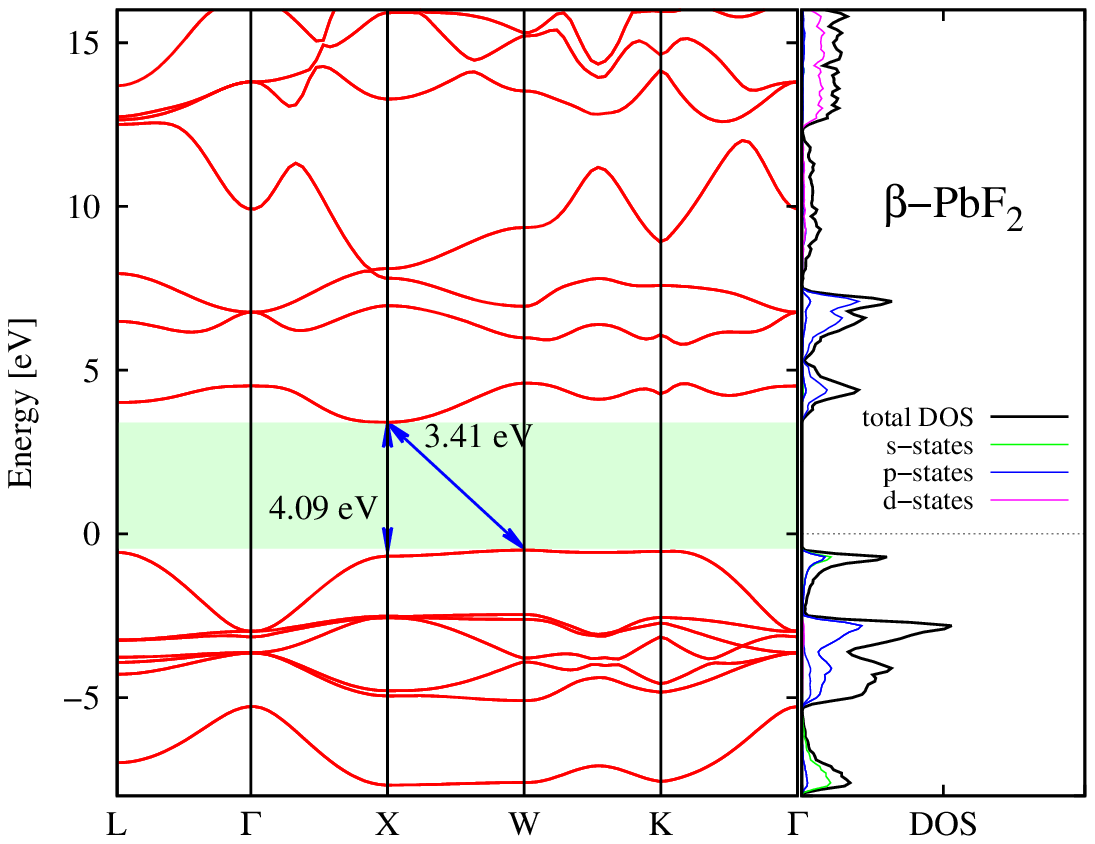}}
\caption{\label{cpt:band graphs}Energy band structures, total and projected density of states (DOS) are shown in the region of the band gap for the IIB-compounds and $\beta $-$PbF_{2}$ .}
\end{figure}

Energy band structure, total and projected density of states (DOS)\cite{note1} are shown in the region of the gap in Figs.~\ref{cpt:band graph},~\ref{cpt:band graphs}. Direct and indirect minimum band gaps are also clearly shown. In Table~\ref{tab:gap bandwidht} these data are shown in comparison to previous results and experimental data; valence bandwidths are also reported in comparison with theoretical and experimental results. In Table ~\ref{tab:mygap} the main vertical transitions are reported for all the compounds.   
In the following lines, in order to extract general trends for the electronic excitation properties in the stable cubic fluorides
under study, we shall compare the results for the different fluorides considering first compounds with cations
belonging to the II group, namely Ca, Sr and Ba, then compounds with metal belonging to IIB group, namely Cd and Hg.
Finally for $\beta $-$PbF_{2}$ data will be analysed separately.
All these compounds show a minimum direct band gap at the $\Gamma$ point, except $\beta $-$PbF_{2}$.
Concerning the smallest gap, the fundamental one, all the compounds herein treated are indirect gap insulators apart from $HgF_{2}$ which shows a direct fundamental band gap (Table~\ref{tab:gap bandwidht}).
Note that absolute values for the gaps show the so-called band-gap underestimate which can be resolved going beyond DFT-LDA,  by using more accurate techniques for the excited states (e.g. GW ones, see Table~\ref{tab:gap bandwidht}). This issue will not be addressed in the present work and will be the subject of future research. Moreover trends on electronic excitation energies as shown in the following should not be affected by the above problem.\cite{cappellini1,cappellini2}
For II-compounds, a decrease of the direct gap at $\Gamma$ of $1.04 ~eV$ is similar to the $1.10~ eV$ decrease after the experimental data (see  also Fig.~\ref{cpt:band graph}). On the other hand, a decrease of $0.54~eV$ for the $X$-$\Gamma$ transitions for calcium and strontium fluorides results which can be compared with a $1.20~eV$ decrease from the experiments.  Considering barium fluorite, the $X$-$\Gamma$ transition is not the minimum, but that which occurs at $(1/4,1/4,0)$-$\Gamma$ (see also Fig.~\ref{fig:band ba}). However, considering the value for the $X$-$\Gamma$ transition shown in Tab.~\ref{tab:mygap}, we confirm a smaller value of $0.31 ~eV$ with respect to the same transition for $SrF_{2}$ (to compare with $0.6~eV$ experimental decrease).
Considering the valence bandwidth, a $1.39 ~eV$ decrease going from $Ca$ to $Ba$ occurs, and that decrease can be compared to a $0.70 ~eV$ experimental one.

For IIB-compounds (see Figs.~\ref{fig:band cd},~\ref{fig:band hg}), the direct gaps at the $\Gamma$ point show  $3.02 ~eV$ difference going from $CdF_{2}$ to $HgF_{2}$.  
While the $CdF_{2}$ presents an indirect fundamental gap between $W\rightarrow\Gamma$, $HgF_{2}$ shows a direct fundamental band gap at $\Gamma$ (see Fig.~\ref{fig:band hg}). Moreover, for the IIB-compounds $HgF_{2}$ presents a larger value of valence bandwidth of $0.59 ~eV$.

For $\beta $-$PbF_{2}$ the minimum direct gap occurs at $X$ instead of at $\Gamma$ as for the other fluorides. It shows an indirect fundamental gap  $ W\rightarrow X $ of $3.41~eV$ (see Fig.~\ref{fig:band pb} and Tab.~\ref{tab:gap bandwidht}). Moreover, the largest value for the valence bandwidth occurs, with a slight overestimate of the experiment ($13\%$).

\begin{table}[tp]
\caption{Energy band gaps (eV) after present work.}
\label{tab:mygap}
\centering
\begin{tabular}{c|cccccc}
\hline
\hline
Direct gap  		& $CaF_{2}$  	& $SrF_{2}$ 	&$BaF_{2}$ 	&  $CdF_{2}$ 	& $HgF_{2}$	& $\beta $-$PbF_{2}$  \\
\hline
$ L\rightarrow L $       & 9.35 	&  10.59 	&9.29 	&    7.62 	          & 4.31   & 4.62 \\
$\Gamma\rightarrow\Gamma$ 	 & 7.71 	&  6.99	&6.67  		& 3.37    & 0.35    & 7.45 \\
$ X\rightarrow X $       & 8.09 	&  8.03   	&7.23 	&   8.18 	          & 5.34    & 4.09 \\
$ W\rightarrow W $       & 8.57 	&  8.44  	&7.68 	&    8.17 	          & 5.45    & 5.10 \\
$ K \rightarrow K $      & 8.50  	&  8.41 	&7.54 	&     8.25 	          & 5.34    & 5.45 \\
\hline
Indirect gap & &  &  &  &  &  \\
\hline
$X\rightarrow\Gamma $      	  &7.43		&   6.89		& 6.79 	&      3.04 	&  ...  &  ...   \\
$L\rightarrow\Gamma $     	  &8.00		&   7.33		& 6.73	&      3.29  	&   ... &  ...   \\
$W\rightarrow\Gamma $     	  &7.54		&   6.97      	  	& 6.90	&      2.94 	&  ...  &  ...    \\
$K\rightarrow\Gamma $     	  &7.55		&   6.98        	& 6.83	&      3.14  	&  ...  &  ...\\
$ (\tfrac{1}{4},~\tfrac{1}{4},~0)\rightarrow\Gamma$ &   ...    	&  ...	& 6.58  &   ... &  ...  & ...   \\
$ \Gamma\rightarrow X $      	  &...		&     ...      	&   ... &       ...    	&  ...  & 6.36\\
$ W\rightarrow X $          	  & ...		&      ...      &   ... &       ...	&  ...  & 3.95 \\
$ K\rightarrow X $          	  & ...		&        ...    &   ... &       ...	&  ...  & 3.98 \\
$ L\rightarrow X $         	  & ...		&      ...     	&   ... &        ...	&  ...  & 4.02\\
\hline
\hline
\end{tabular}
\end{table}

With respect to direct transitions, considering the data in Tab.~\ref{tab:mygap}, all II-compounds show a decreasing trend at all main symmetry points, except for $L$. At that point $SrF_{2}$ shows  a $1.24 ~eV$ ~larger band gap with respect to $CaF_{2}$ and a $1.3 ~eV$  ~larger value with  respect to $BaF_{2}$.
For IIB-compounds, the direct transitions show a larger difference (up to $3.31 ~eV$) going from $Cd$ to $Hg$.

To complete the picture, the indirect gaps between the top of the valence band at high symmetry points  to the bottom of the conduction bands ($\Gamma$ for II- and IIB-compounds, $X$ for $\beta$-$PbF_{2}$) are shown in the second part of Tab.~\ref{tab:mygap} .

\section{CONCLUSIONS}
\label{sec:conclusion}
The DFT-LDA electronic structures for ground-state and excited states for some cubic fluorides have been studied in detail. The electronic density of states (DOS) at the gap region for all the compounds and their energy-band structure have been calculated and compared with existing experiments and previous theoretical results. The electronic energy-bands and transition energies are given and discussed. Within the same DFT-LDA scheme, general trends for the ground-state parameters and the DOS are also given.
The above trends show good comparison with experimental data and theoretical results.
Relatively to electronic excitations, 
the conduction bands for II-compounds are mostly dominated by the cations d-orbitals, while for the IIB-compounds the tail of the DOS at the lowest conduction bands shows a largely s-type character. 
The obtained DFT-LDA  valence bandwidths agree with experimental values within $~30 \%.$
The present systematic DFT-LDA study is of particular interest for future researches on excited states and optical properties calculations of fluorides. We plan to carry out such calculations in next future by using those techniques particularly devoted to that issue.

The authors acknowledge computational support provided by COSMOLAB (Cagliari, Italy) and CASPUR (Rome, Italy).
Discussions with  V. Fiorentini are gratefully acknowledged.

\bibliographystyle{plain}

\begin{thebibliography}{99}
\bibitem{note1}
Electronic density of states (DOS) have been  calculated using a Brillouin zone sampling with an $8\times 8\times 8$ Monkhort-Park grid and  Gaussian smearing of width 0.1 eV.
\end{thebibliography}

\end{document}